\documentclass[aps,pra,twocolumn,secnumarabic,groupedaddress,nobalancelastpage,
amsmath,amssymb]{revtex4}

\usepackage{graphics}		
\usepackage{graphicx}		
\usepackage{bm}			
\usepackage{mathptmx}		

\newcommand{\eq}[1]{Eq.~(\ref{eq:#1})}		
\newcommand{\ffig}[1]{Figure~\ref{fig:#1}}	
\newcommand{\fig}[1]{Fig.~\ref{fig:#1}}		
\newcommand{\ssect}[1]{Section~\ref{sec:#1}}	
\newcommand{\sect}[1]{Sec.~\ref{sec:#1}}	

\begin{document}

\title{Experimental investigation of planar ion traps}

\author{C. E. Pearson}
\author{D. R. Leibrandt}
\author{W. S. Bakr}
\author{W. J. Mallard}
\author{K. R. Brown}
\author{I. L. Chuang}
\affiliation{Center for Bits and Atoms {\em and} Department of
Physics, Massachusetts Institute of Technology, Cambridge,
Massachusetts 02139, USA} 

\date{\today}

\begin{abstract}

Chiaverini et al.~[Quant.~Inf.~Comput.~\textbf{5}, 419 (2005)] recently
suggested a linear Paul trap geometry for ion trap quantum computation
that places all of the electrodes in a plane. Such planar ion traps are
compatible with modern semiconductor fabrication techniques and can be
scaled to make compact, many zone traps.  In this paper we present an
experimental realization of planar ion traps using electrodes on a
printed circuit board to trap linear chains of tens of 0.44 $\mu$m
diameter charged particles in a vacuum of 15 Pa ($10^{-1}$ torr). With
these traps we address concerns about the low trap depth of planar ion
traps and develop control electrode layouts for moving ions between trap
zones without facing some of the technical difficulties involved in an
atomic ion trap experiment. Specifically, we use a trap with 36 zones
(77 electrodes) arranged in a cross to demonstrate loading from a
traditional four rod linear Paul trap, linear ion movement, splitting
and joining of ion chains, and movement of ions through intersections. 
We further propose an additional DC biased electrode above the trap
which increases the trap depth dramatically, and a novel planar ion trap
geometry that generates a two dimensional lattice of point Paul traps.

\end{abstract}

\maketitle

\section{Introduction}

In recent years, the quantum computing community has demonstrated the
basic building blocks of a scalable ion trap quantum computer
\cite{cirac95,wineland98,kielpinski02,steane05}.  The design uses the
electronic states of ions trapped in a radiofrequency (RF) Paul trap as
qubits and accomplishes logic gates and state readout by laser-ion
interactions.  Ions are shuttled between zones in an array of ion traps
to perform two qubit gates between arbitrary pairs of qubits. Recent
experiments have accomplished state readout \cite{blatt88}, one and
two qubit gates \cite{nagerl99,turchette98}, and ion shuttling in
straight lines and through tees \cite{rowe02,hensinger05}.  Scaling this
architecture up to the many thousands of qubits necessary for a useful
computation, however, will involve significant physics and engineering
challenges. 

\begin{figure}
\includegraphics[width=8.2cm]{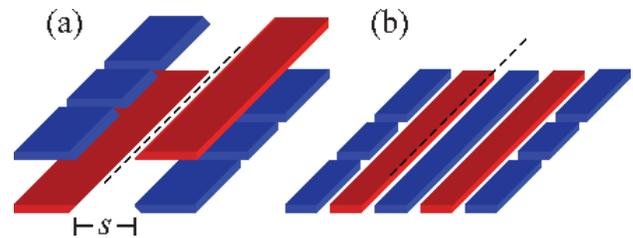} \caption{Schematic illustrations of (a) a two-level linear RF Paul trap
representative of what is currently used in quantum computation experiments,
and (b) the five electrode planar ion trap suggested by Chiaverini et
al.~\cite{chiaverini05}.  Ions are trapped along the trap axis, shown as a
dotted line.  An RF potential is applied to the red electrodes to provide
radial confinement, and DC potentials are applied to the blue control
electrodes to provide axial confinement and to shuttle ions along the trap
axis.  Typical dimensions for current two-level traps are a slot width $s$
of 200--400 $\mu$m.} \label{fig:trap-schematics}
\end{figure}

One of these challenges is to build a many zone ion trap array capable of
holding several thousands of ions in memory zones and moving arbitrary pairs
of them together in interaction zones.  The ion traps used for quantum
computation are based on the linear RF Paul trap
\cite{prestage89,paul90,raizen92,ghosh95}. Current experiments typically use
gold electrodes deposited on two or more alumina substrates with a geometry
similar to that shown in \fig{trap-schematics}(a)
\cite{turchette00,deslauriers04}.  Ions are confined radially to the null
axis of an RF quadrupole electric field, while axial confinement and ion
movement operations are accomplished by DC electric fields. While these
multi-level traps can be adapted to microfabrication techniques
\cite{madsen04}, it is not clear whether they can be scaled to many zone
traps because they require slots through the trap structure.  This will make
trap topologies that include loops difficult, as there will be islands of
electrodes that will have to be mechanically supported and electrically
connected.

Chiaverini et al.~\cite{chiaverini05} proposed using a planar RF Paul
trap geometry for ion trap quantum computing which is easy to scale up to
many zone traps and amenable to modern microfabrication techniques.  The
electrodes all lie in a plane and ions are trapped above the plane of
the electrodes \cite{janik90}.  In the five electrode planar trap design
shown in \fig{trap-schematics}(b), the center and outermost electrodes
are held at RF ground while the remaining two electrodes are biased with
an RF potential for radial confinement. Either the center electrode or
the outermost two electrodes can be segmented and DC biased for axial
confinement.  Such planar ion traps can be built using silicon VLSI
technology and thus have the capability to scale to arbitrarily large
and complex trap arrays \cite{kim05a}.

Several challenges will have to be addressed, however, before planar ion
traps can be used for quantum computing.  Perhaps the most significant
challenge is that planar ion traps have trap depths that are only of
order 1\% that of a multi-level trap of comparable dimensions
\cite{chiaverini05}. While this is not a problem once the ions are
loaded and laser cooled to near the ground state of the motion, it makes
loading from a thermal ion source at room temperature or above
difficult.  A second challenge is to show that it is possible to perform
the three basic ion movement operations required to implement the
Kielpinski et al.~\cite{kielpinski02} ion trap quantum computer
architecture in a planar trap: ballistic ion transport in straight
lines, splitting and joining pairs of ions in a single trap, and
shuttling ions around corners in intersections.  Further challenges are
that optical access for laser cooling is blocked along the axis
orthogonal to the trap substrate, and that ion heating is enhanced first
by the close proximity of the electrodes that is required to achieve
a reasonable trap depth, and second by the more resistive semiconductor
materials used for VLSI \cite{yurke05}.

In this work we address the first and second challenges.  The latter
challenges are addressed elsewhere \cite{kim05a,yurke05}.  We
investigate planar ion traps using printed circuit board (PCB)
electrodes to trap macroscopic charged particles in a rough vacuum
environment.  We do not cool our ions, and imaging is accomplished by
classical (off resonant) scattering of laser light.  Because these
experiments do not require an ultra high vacuum environment, we are able
to achieve a fast cycle time for testing different electrode layouts. 
We optimize the electrode geometry to maximize the trap depth, and
further demonstrate loading from a conventional four rod linear Paul
trap acting as a reservoir of ions. Atomic ions could be laser cooled in
the four rod trap before being transferred to the planar trap to
circumvent the problem of low trap depth.  We test trap layouts with
segmented center electrodes and segmented outer electrodes, and have
accomplished all three of the movement operations required for the
Kielpinski et al.~\cite{kielpinski02} architecture.  While transport,
splitting and joining \cite{rowe02}, and movement through intersections
\cite{hensinger05} have already been achieved in multi-level traps, this
work represents the first realization of these operations in planar ion
traps.

This paper proceeds as follows.  \ssect{design} offers a general discussion
of planar ion trap design and a detailed description of our experimental
setup.  In \sect{characterization}, we characterize the macroscopic charged
particles we use to test the traps.  \ssect{performance} presents the
results of our investigation of planar ion traps including secular motion
and ion movement experiments.  In \sect{variations} we discuss some of the
various alternative trap geometries we have tested and propose a novel
planar ion trap geometry that forms a lattice of point RF Paul traps.
Finally, in \sect{conclusions} we summarize and suggest how to proceed in
experimentally demonstrating atomic planar ion traps.

\section{Experimental design}
\label{sec:design}

The design of a planar trap for trapping macroscopic particles involves
three main considerations: the ion source, the ion loading scheme, and
the PCB design. Particle ion sources in common use include electrospray
ionization \cite{ESI:1}, laser ablation \cite{MALDI:1, MALDI:2}, and
piezoelectric particle generation \cite{Piezoelectric:1}. The careful
choice of an ion loading mechanism is important because of the low trap
depth of planar traps. The planar trap described in this paper is loaded
using a four rod trap with a higher trap depth. Designing the PCB
involves choosing the electrode dimensions to get the desired trap
depth, ion height above the substrate, and secular frequencies; choosing
the layout of the control electrodes to allow for performing the desired
movement operations; and choosing a PCB substrate that has a high
voltage breakdown threshold, does not accumulate charge easily, and is
vacuum compatible in the case of a trap in a UHV environment. This
section begins by presenting some general considerations for planar trap
design and proceeds to describe the experimental setup.

\subsection{Trap design}
\label{sec:trap-design}

The dynamics of an ion in a linear RF Paul trap are determined by solving
the classical equations of motion \cite{ghosh95}.  For an RF quadrupole
electric potential
\begin{equation}
\phi(x,y,t) = \frac{x^2 - y^2}{2 r_0^2} \left( U - V \cos(\Omega t)
\right) \ ,
\end{equation}
the $x$ and $y$ equations of motion take the form of Mathieu equations:
\begin{equation}
\label{eq:mathieu-x}
\frac{d^2 x}{d \tau^2} + \left( a - 2 q \cos(2 \tau) \right) x = 0
\end{equation}
and
\begin{equation}
\frac{d^2 y}{d \tau^2} - \left( a - 2 q \cos(2 \tau) \right) y = 0 \ .
\end{equation}
Here $\tau = \Omega t/2$ is a dimensionless time, the Mathieu parameters
$q = (2 Q V)/(m r_0^2 \Omega^2)$ and $a = (4 Q U)/(m r_0^2 \Omega^2)$
are dimensionless RF and DC voltages, and $Q$ and $m$ are the ion charge
and mass.  The motion is stable (i.e., the components of the ion
position vector $x$ and $y$ are bounded in time) in regions of $a-q$ parameter
space.  In particular, for $a = 0$ the motion is stable for $0 < q \le
q_{max} = 0.908$.

\begin{figure}
\includegraphics[width=8.2cm]{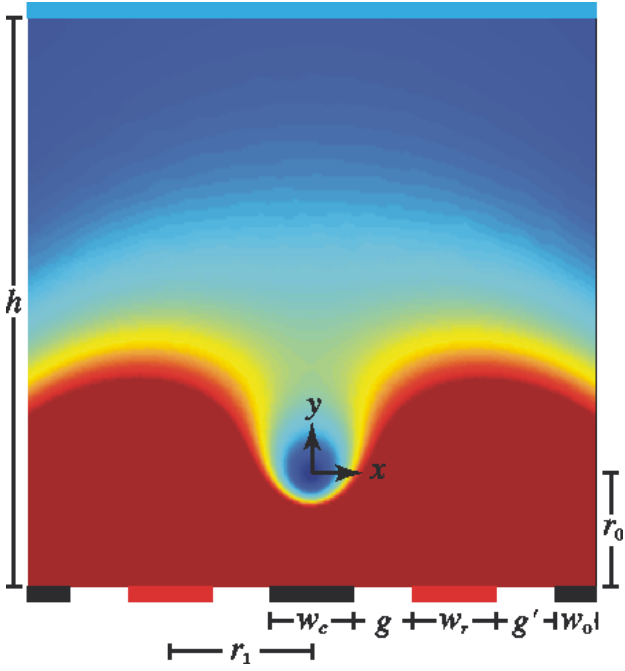}
\caption{\label{fig:secular-potential}Secular potential for a
planar ion trap.  The particular geometry depicted is that of the cross trap
described in \sect{expt-apparatus}. The secular potential is plotted on a
linear color scale as a function of $x$ and $y$, with blue representing the
lowest and red representing the highest secular potential values.  Secular
potential values above twice the trap depth are truncated for clarity.  The
ion is located at the origin of the coordinate system.  An RF potential $V
\cos(\Omega t)$ is applied to the red electrodes, a DC potential $U$ is
applied to the blue top plate electrode ($U = 0$ in this figure), and the black electrodes are
grounded.}
\end{figure}

In the pseudopotential approximation, where $q \ll 1$, the ion motion
along axis $i$ can be decomposed into slow, large amplitude secular
motion at the secular frequency $\omega_i$ and fast, small amplitude
micromotion at the RF drive frequency $\Omega$ \cite{dehmelt67}.  For an
arbitrary electric potential of the form
\begin{equation}
\phi(x,y,z,t) = \phi_{RF}(x,y,z) \cos(\Omega t) + \phi_{DC}(x,y,z) \ ,
\end{equation}
the secular motion is determined by a secular potential
\begin{equation}
\psi_{sec}(x,y,z) = \frac{Q^2}{4 m \Omega^2} | \bm{\nabla} \phi_{RF}(x,y,z) |^2 + Q \phi_{DC}(x,y,z) \ .
\end{equation}
Using the secular potential we can calculate the height $r_0$ of the ion
above the plane of the electrodes, the secular frequencies
$\omega_i$ in the harmonic region of the potential, and the trap depth 
$\psi_{sec,0}$. The trap depth is defined in the usual way as the
energy an ion would need to escape the trap.  The secular potential for
the planar trap used in these experiments is shown in \fig{secular-potential}.

\begin{figure}
\includegraphics[width=8.2cm]{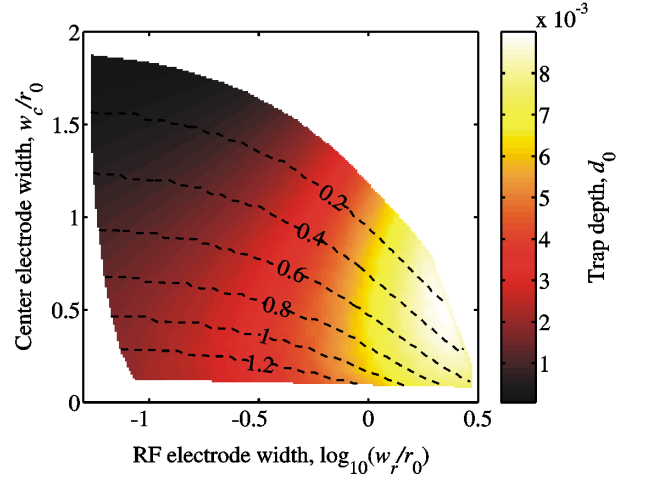}
\caption{\label{fig:trap-depth-vs-geometry}Trap depth versus electrode
widths.  The normalized trap depth $d_0$ is plotted in color as a function
of $\log_{10}(w_r/r_0)$ and $w_c/r_0$ for $h/r_1 = 706$.  The dashed lines are
contours of constant $g/r_0$.  Note that the maximum trap depth is obtained
at $w_c/r_0 \approx 0.5$ and $w_r/r_0$ as large as possible.}
\end{figure}

\begin{figure}
\includegraphics[width=8.2cm]{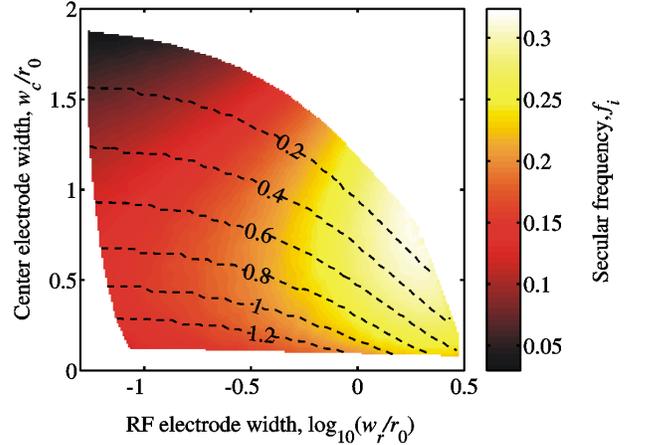}
\caption{\label{fig:frequency-vs-geometry}Secular frequency
versus electrode widths.  The secular frequency at zero ion temperature
$f_i$ is plotted in color as a function of $\log_{10}(w_r/r_0)$ and
$w_c/r_0$ for $h/r_1 = 706$.  The dashed lines are contours of constant
$g/r_0$.}
\end{figure}

In order to determine the optimum relative sizes of the trap electrodes, we
have calculated the secular potential numerically for the planar trap
geometry defined in \fig{secular-potential} over a range of dimensions using
a two dimensional finite element electrostatic package named BELA
\cite{meeker04}.  Here $w_c$ is the width of the center electrode, $w_r$ is
the width of the RF electrodes, $w_o$ is the width of the outer electrodes,
$g$ is the width of the gap between the center and the RF electrodes, $g'$
is the width of the gap between the RF and outer electrodes, $h$ is the
height of a planar electrode which is parallel to and above the trap
electrodes, $r_0$ is the height of the ion above the trap electrodes, and
$r_1 = (w_c + w_r)/2 + g$ is a measure of the trap size.  We set $g' = g$
and $w_o \to \infty$, and varied $w_c$, $w_r$, and $h$.  Note that $h \to
\infty$ for a true planar trap. Figures \ref{fig:trap-depth-vs-geometry} and
\ref{fig:frequency-vs-geometry} show the normalized trap depth $d_0$ defined
by
\begin{equation}
\psi_{sec,0} = \frac{Q^2 V^2}{4 m r_0^2 \Omega^2} d_0 \ ,
\end{equation}
and the normalized secular frequency
$f_i$ defined by
\begin{equation}
\label{eq:secularfreq}
\omega_i = \frac{Q V}{\sqrt{2} m r_0^2 \Omega} f_i
\end{equation}
as functions of $w_c/r_0$ and $w_r/r_0$ for $h/r_1 = 706$.  Note that $d_0 =
f_i = 1$ for a perfect quadrupole trap and that we consider the secular
frequency in the harmonic region of the potential where it is the same along
both axes (this is appropriate when the ions are at low temperature).

Using Figs. \ref{fig:trap-depth-vs-geometry} and
\ref{fig:frequency-vs-geometry}, we can design a planar trap to maximize
the trap depth subject to the experimental constraints.  As an example,
consider a planar ion trap for $^{88}$Sr$^+$.  The ion height $r_0$ is set
by either the acceptable ion heating rate or by how tightly the cooling
and other lasers can be focused such that they do not scatter off the
surface of the trap. Suppose we pick $r_0 = 500~\mu$m.  The electrode
widths $w_c$ and $w_r$ should be chosen to maximize the trap depth using
\fig{trap-depth-vs-geometry}, subject to the constraint that the gap
between them $g$ must be large enough that the RF voltage does not
induce electrical breakdown.  In general, the optimum center electrode
width is around $0.5 r_0$, and the optimum RF electrode width is as wide
as possible given practical considerations such as the capacitance of a
large electrode.  For our example, suppose further that $g = 0.5 r_0 =
250~\mu$m and $w_r = r_0 = 500~\mu$m.  Figs.~
\ref{fig:trap-depth-vs-geometry} and \ref{fig:frequency-vs-geometry},
then, give $w_c = 0.58 r_0 = 290~\mu$m, $d_0 = 0.0065$, and $f_i =
0.28$. At $V = 500$ V and $\Omega = 2 \pi \times 10$ MHz, this
corresponds to $\psi_{sec,0} = 0.47$ eV and $\omega_i = 2 \pi \times
1.1$ MHz.

\begin{figure}
\includegraphics[width=8.2cm]{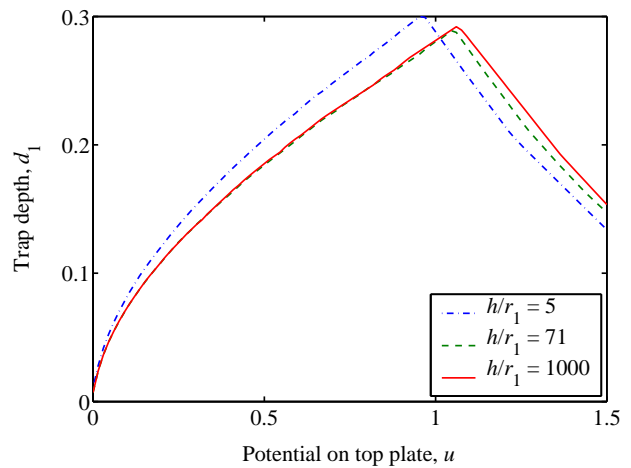}
\caption{\label{fig:trap-depth-vs-top-plate}Trap depth versus
top plate location and potential.  The normalized trap depth $d_1$ is
plotted as a function of $u$ for $w_c/r_1 = w_r/r_1 = 0.6$ and several
values of $h/r_1$.  For $u \lesssim 0.2$, the ion escapes straight up in the
positive $y$ direction; for $0.2 \lesssim u \lesssim 1$, the ion escapes out
to the side; and for $1 \lesssim u$ the ion escapes straight down in the
negative $y$ direction.  Note that the maximum trap depth $d_1 \approx 0.29$
is obtained at $u \approx 1$.}
\end{figure}

We have found that by adding a planar top electrode (i.e., choosing $h$ to
be finite) biased at a positive DC potential $U$ we can increase the trap
depth by up to a factor of a 40 for a given RF amplitude and frequency.
\ffig{trap-depth-vs-top-plate} shows the normalized trap depth defined by
\begin{equation} \psi_{sec,0} = \frac{Q^2 V^2}{4 m r_1^2 \Omega^2} d_1 \ ,
\end{equation} as a function of the normalized voltage on the top plate,
\begin{equation} u = \frac{4 m r_1^2 \Omega^2}{Q V^2} \frac{r_1}{h} U \ ,
\end{equation} for $w_c/r_1 = w_r/r_1 = 0.6$ and several values of $h/r_1$.
Note that the trap depth is normalized differently for this plot than it was
for \fig{trap-depth-vs-geometry}. This is because $r_0$ is a logical
starting point for selecting the electrode dimensions, but $r_1$ is the
fixed length scale once the trap is built. At $u = 0$, the smallest barrier
in the secular potential over which the ion can escape is in the positive
$y$ direction.  As $u$ increases, the minimum energy escape path shifts
first to the sides and then to the negative $y$ direction. For a given
geometry, the trap depth increases with $u$ until the minimum energy escape
path is straight down to the center electrode, then decreases with a further
increase in $u$.  The maximum trap depth occurs at $u \approx 1$ and is
given by $d_1 \approx 0.29$, or a little more than one fourth the trap depth
of a standard four rod ion trap with similar dimensions. The secular
potential depends only weakly on $h/r_1$ because the effect of nonzero $u$
is well approximated by a constant electric field in the regime $h/r_1 \gg
1$ that we consider.  Note that because the DC potential pushes the ion off
of the RF quadrupole null, there will be increased micromotion.  To first
order in a harmonic potential, the micromotion amplitude is
\begin{equation}
\label{eq:micromotion}
q \Delta y / 2
\end{equation}
where $\Delta y$ is the displacement of the ion from the RF null.  In
practice we envision turning the DC potential on for loading, then gradually
turning it off as the ions are laser cooled. Continuing our previous
example, with a top plate electrode $h = 1$ cm above the trap biased at $U =
100$ V, we get a trap depth $\psi_{sec,0} = 4.6$ eV and a micromotion
amplitude of 15 $\mu$m using \eq{micromotion}.  A direct numerical
calculation of the classical ion trajectory along the $y$-axis predicts a
micromotion amplitude of 20 $\mu$m, confirming that \eq{micromotion} is a
reasonable approximation despite the rather non-harmonic nature of the
potential.

\subsection{Experimental Apparatus}
\label{sec:expt-apparatus}

In this section, we discuss the specifics of the trap's design and
construction, our source of ions, the method for loading the planar
trap, the environment used for performing our experiments, and our
methods for measurement and control of ions.

\begin{figure}
\includegraphics[width=0.45\textwidth]{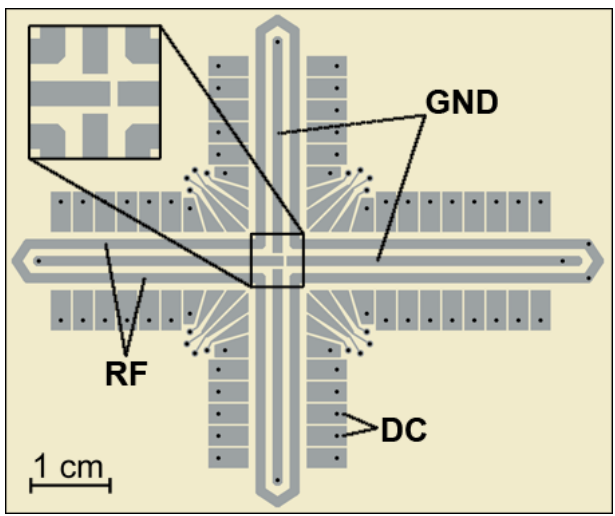} \caption{Top
view of the trap.  Opposing electrodes in straight sections of the trap are
electrically connected, but near the intersection both the outer and center
electrodes are separated to provide finer control.  Electrical connections
are made via surface mount headers on the underside.  The RF loop at each
end helps prevent ions from leaking out axially.} \label{fig:cross}
\end{figure}

The planar trap we investigated is made up of four straight arms joined
at a cross intersection (\fig{cross}).  It is a printed circuit board
(PCB), made using standard techniques.  The electrodes are tin coated
copper, and the substrate is GML-1000, a microwave laminate.  The trap
is a five electrode design, with outer electrodes segmented to control
the axial potential.  Connections to these electrodes are made on the
bottom layer of the PCB using surface mount headers.  Most opposing
pairs of control electrodes are electrically connected, but the two
pairs nearest the intersection on each arm are electrically independent
for finer control.  The center electrodes are also segmented at the
intersection to provide additional control in that region. The middle
three electrodes are all 1.27 mm wide, the electrode spacing is 0.89 mm,
and the outer control electrodes are 2.5 mm long.  We do not use the
optimal electrode sizes as discussed in \sect{trap-design} due to the
practical considerations of printed circuit board manufacture,
specifically minimum feature and drill sizes.

\begin{figure}
\includegraphics[width=0.45\textwidth]{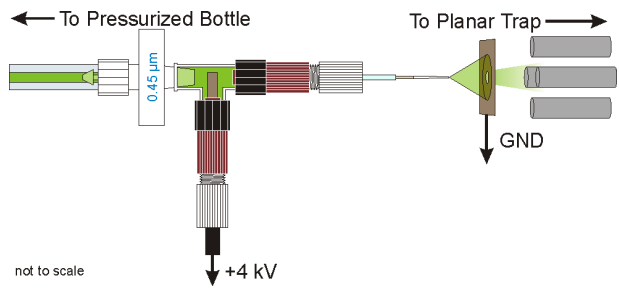} \caption{The
ESI apparatus.  Microspheres suspended in a mixture of methanol and water
are forced from a pressurized bottle, through a 0.45 $\mu$m filter and
past a copper wire with +4 kV applied voltage, and out the capillary tip.
The charged spray passes through a mask and into a four rod trap.}
\label{fig:electrospray}
\end{figure}

Electrospray ionization (ESI) provides ions to the trap.  This technique,
commonly used with linear quadrupole filters for mass spectroscopy, applies
high voltage to a liquid solution at a sharp tip.  Strong electric fields at
the tip blow off fine droplets of solution, and as solvent evaporates from
these charged droplets, self-repulsion breaks them up into smaller
particles, eventually producing individual charged solute particles.
Following \cite{Cai:1}, we begin with a 5\% suspension of 0.44 $\mu$m
diameter aminopolystyrene spheres.  We then prepare a solution buffered to
pH = 3.9, add methanol to produce a 4:1 methanol/solution mixture, and add
the spheres to produce a 0.05\% suspension.  The suspension is then placed
in a sealed bottle, and pressurized to 50-100 kPa.  This forces the liquid
through a 0.45 $\mu$m filter to block any clusters of microspheres, and out
the electrospray tip.  The tip itself is made from fused silica capillary
tubing, which is heated, stretched to produce a neck, and cleaved to produce
a 100 $\mu$m opening.  We apply 4 kV directly to the liquid with a copper
wire inserted in the fluid near the tip (\fig{electrospray}).  The spray
plume from the tip is directed through a grounded mask and into the end of a
four rod trap.

\begin{figure}
\includegraphics[width=0.45\textwidth]{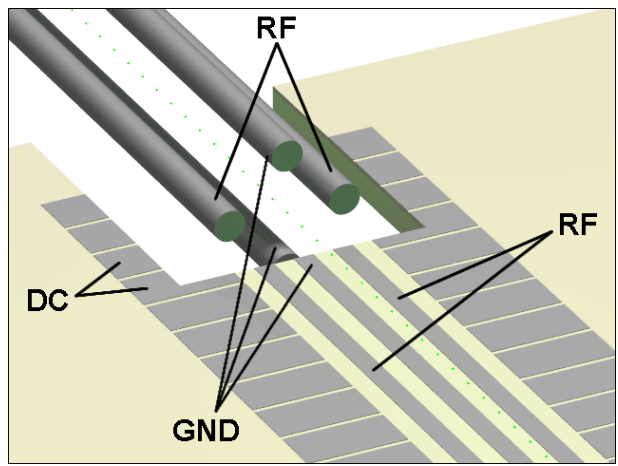}
\caption{Four rod to planar interface.  A slot cut in the planar trap
allows for a common trap axis.  The large particles form a linear Wigner
crystal immediately and are pushed into the planar trap.}
\label{fig:4rodToPlanar}
\end{figure}

By arranging the rods so that they match the strips of a planar trap, we
interface the four rod trap directly to the planar trap.  A slot cut in
the side of the PCB allows us to lower the four rod trap to a point
where its trap axis coincides with that of the planar trap
(\fig{4rodToPlanar}).  Mutual repulsion of ions loaded in the far end of
the four-rod trap forces them along, eventually pushing them into the
planar trap.  When completely full, the planar trap can hold $\sim$50
ions.

\begin{figure}
\includegraphics[width=0.45\textwidth]{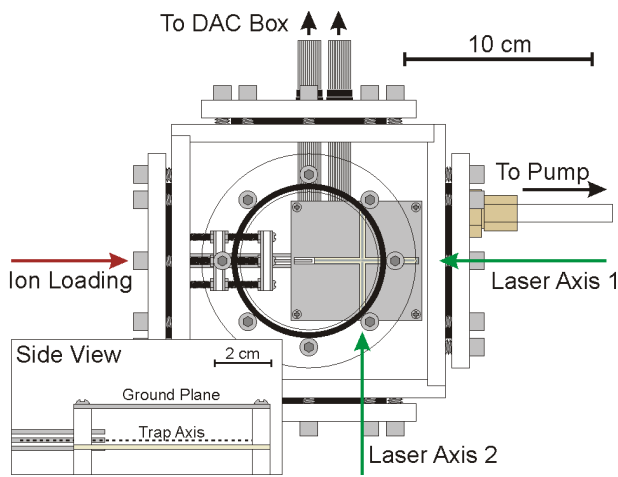}
\caption{Top view of the vacuum chamber.  Ions are loaded into the left
side, and then a flange is screwed into place and the air evacuated
slowly.  Two green lasers illuminate the trap axes.}
\label{fig:acrylicBox}
\end{figure}

Electrospray ionization must be performed at or near atmospheric pressure,
and experiments on free trapped particles require a vacuum, so we enclose
both four rod and planar traps in a custom built clear acrylic box
(\fig{acrylicBox}), and load through an open side of the box.  We perform
all loading at atmospheric pressure, then seal the box shut and carefully
pump down.  We use a needle valve to control flow rate during pumping, since
any significant air flow can push ions out of the trap.  The lowest
attainable pressure in this enclosure is about 15 Pa ($10^{-1}$ torr).
Green (532 nm) laser light directed along the trap axes illuminates ions by
classical (off resonant) scattering.  The ions are clearly visible by eye,
and we photograph them using an ordinary camera lens mounted on a CCD
camera.

Also shown in \fig{acrylicBox} is a metal plate installed above the
trap to increase its trap depth and shield from stray static
charge.  Machined slots in this plate allow us to view ions from above,
but we can also see them from the side.  This slotted top plate design
has the advantage of masking laser scatter, but for more complex trap
topologies it might be easier to use a transparent conductor such as
indium tin oxide or a thin film of gold
\cite{gordon00,granqvist02,sawada01}.  This could be deposited on a
glass plate or directly on the vacuum window or an imaging optic inside
the vacuum chamber.

\begin{figure}
\includegraphics[width=0.45\textwidth]{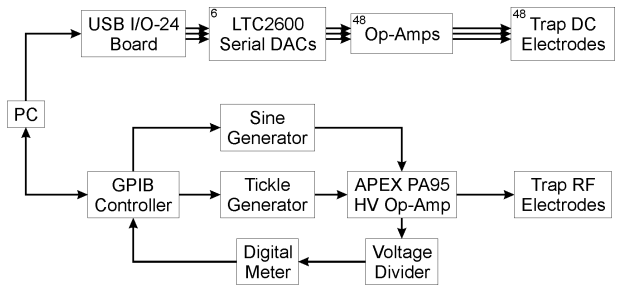}
\caption{Control electronics for the planar trap.}
\label{fig:electronics}
\end{figure}

The electronics used to generate the RF and DC potentials for the trap are
shown in \fig{electronics}.  Frequencies for which the microspheres are
stable range from a few hundred hertz to several kilohertz, depending on the
pressure.  This is in the audio frequency (AF) range, but we will continue
to refer to it as RF\@.  Since these frequencies are easily synthesized, we
use a function generator as the source, then use an active high voltage
amplifier to reach the target voltage range.  A small tickle signal can be
added to the main RF signal before amplification to probe for resonances in
ion motion, allowing us to determine secular frequencies (\sect{secular}).
In this design, the upper limit to RF voltage is set by arcing between RF
and ground electrodes, which occurs at about 400 V amplitude.  Typical
operating conditions are 250 V and 1.5 kHz.

To provide DC control of the trap's 48 independent electrodes, we
use a board with 24 digital output channels to address six serial octal
digital to analog converter (DAC) integrated circuits.  The outputs of
these DAC chips are then amplified to $\pm$20 V, sufficient to axially
confine or to move these particles.

\section{Ion characterization}
\label{sec:characterization}

Our goal in this work is to study the properties of planar traps rather
than the ions loaded into them. To be able to draw conclusions about
atomic ion traps from work on macroscopic ions, we must (i) scale away
the charge-to-mass dependence from measured parameters and (ii)
understand the effect of damping on macroscopic ions since it is
negligible for atomic ions.  \ssect{eoverm} discusses measurement of the
$Q/m$ spectrum of the ions. \ssect{damping} discusses the effects of air
drag on the ions' stability and motion.

\subsection{\label{sec:eoverm}Charge to mass ratio}

The principal drawback of using macroscopic charged particles to
investigate a trap design is that, unlike atomic ions, these particles
are not identical.  Electrospray ionization (ESI) is a soft ionization
technique that produces multiply charged ions with a significant spread
in the charge state. The variation in the mass of the microspheres
($\approx$ 10\%) also contributes to broadening the charge-to-mass
distribution. Nevertheless, we are able to keep the $Q/m$ spectrum
constant by maintaining constant electrospray pressure, voltage,
capillary diameter and loading parameters.

\begin{figure}[t]
\includegraphics[width=8.2cm]{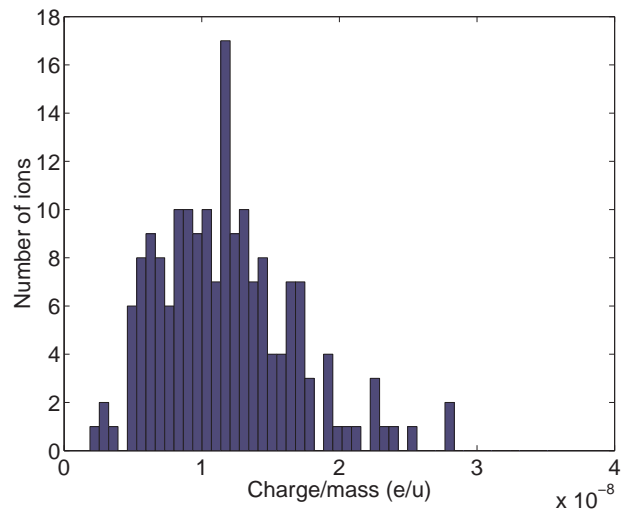} \caption{Charge-to-mass
distribution of ESI-generated ions. The mean $Q/m$ is $1.17 \times 10^{-8}$
$e$/amu and the standard deviation is $0.49 \times 10^{-8}$ $e$/amu.}
\label{fig:eoverm}
\end{figure}

To measure the $Q/m$ spectrum of the ions, groups of about
five ions at a time were loaded into the four rod trap with $\Omega
= 2\pi \times 2$ kHz and $V = 250$ V. With the chamber pressure at
70 Pa, the RF frequency was lowered while observing the ejection
frequency $\Omega_{ej}$ of each ion.  The charge-to-mass ratio was
calculated using the formula $Q/m = q_{max} r_0^2
\Omega_{ej}^2/(2V)$, where $q_{max} = 0.908$, the value of the
Mathieu $q$ parameter at the boundary of the first stability region of a
quadrupole trap.  \ffig{eoverm} shows a spectrum obtained for 178
ions with mean $Q/m = 1.17 \times 10^{-8} e$/amu (electronic charges per
nucleon mass).

\subsection{\label{sec:damping}Damping of ion motion by background gas}

When trapping macroscopic ions at atmospheric pressure or in rough
vacuum, the background gas exerts a drag force on the ions that both
stabilizes the radial motion and slows down the axial ion movement
operations.  It is important to verify that the background pressure in
our experiments is low enough to make accurate assessments of the trap
performance.

For small Reynolds number $N_{Re} = 2 \rho \dot{x} R/\mu$ and small 
Knudsen
number $N_{Kn} = \lambda/R$, the drag on a sphere is given by Stokes' 
law
\begin{equation}
\label{eq:stokes}
F_D = -6 \pi \mu R \dot{x} \ .
\end{equation}
Here $\rho$ is the density of the gas, $R$ is the radius of the sphere,
$\mu$ is the dynamic viscosity, $\lambda$ is the mean free path, and
$\dot{x}$ is the ion velocity along the $x$ axis.
When the Knudsen number is large, Stokes' law overestimates the drag
because the flow is no longer continuous.  We divide \eq{stokes} by an
empirical slip correction factor
\begin{equation}
C(N_{Kn}) = 1 + N_{Kn} \left( 1.165 + 0.483 e^{-0.997/N_{Kn}} \right)
\end{equation}
to obtain an expression for the drag force which is valid at large
Knudsen number \cite{kim05b}.

\begin{figure}
\includegraphics[width=8.2cm]{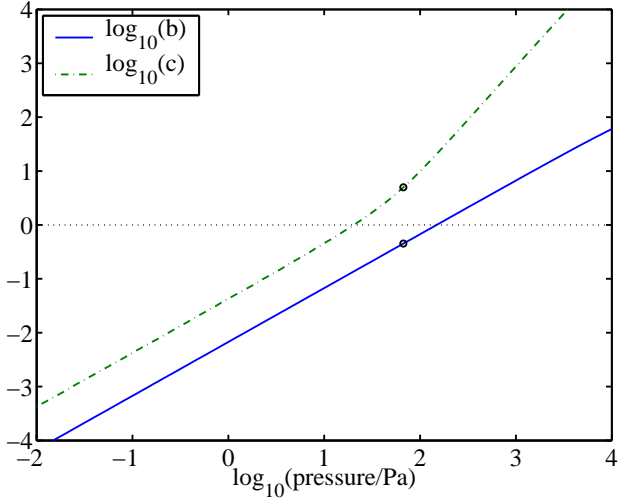}
\caption{\label{fig:drag-parameters}Drag coefficients describing ion
stability and the speed of linear ion shuttling versus background pressure.
The dimensionless drag coefficients $b$ and $c$ for stability and ion
shuttling, respectively, are plotted as functions of the vacuum chamber
pressure on a log-log scale.  The drag coefficients are computed using $R$ =
0.22 $\mu$m, $Q = 5.3 \times 10^{-17}$ C, $m = 4.7 \times 10^{-17}$ kg, $\mu
= 1.83 \times 10^{-5}$ kg m$^{-1}$ s$^{-1}$, $\lambda = (\textrm{67.3 nm})
(10^5\textrm{ Pa}/p)$, $\Omega = 2 \pi \times 5$ kHz, $d = 1$ mm, and $E =
(\textrm{1 V})/(\textrm{1 mm})$; which is appropriate for the cross trap
assuming that we are trapping single microspheres. The circles indicate the
values of the drag coefficients at the pressure used in our experiment.}
\end{figure}

We determine the effect of drag on stability by considering a modified
dimensionless equation of motion
\begin{equation}
\frac{d^2 x}{d \tau^2} + b \frac{d x}{d \tau} + \left( a - 2 q \cos(2
\tau) \right) x = 0
\end{equation}
where the dimensionless drag coefficient
\begin{equation}
b = \frac{12 \pi \eta R}{C(N_{Kn}) m \Omega}.
\end{equation}
A similar equation holds for the $y$ motion.  For $b = 0$, this reduces to
\eq{mathieu-x}.  Numerical computations verify that the region of stable
trapping in $a-q$ parameter space grows with increasing $b$
\cite{winter91,nasse01}.  In particular, the maximum stable Mathieu $q$
parameter at $a = 0$ goes from $q_{max} = 0.908$ at $b = 0$ to $q_{max} =
1.05$ at $b = 0.45$.  The drag parameter $b$ is plotted as a function of
vacuum pressure $p$ in \fig{drag-parameters}.  At 70 Pa (0.5 torr), $b =
0.45$ is already less than 1 so air drag has only a small effect on the trap
stability in these experiments.

The speed of ion motion in our shuttling experiments, however, is
significantly affected by drag.  The physical reason for this is that the
characteristic time for shuttling experiments is much longer than the
characteristic time for stability ($1/\Omega$).  As a simple model, consider
an ion which starts from rest at $z = 0$ in an axial potential
\begin{equation}
\phi(z) = \left\{ \begin{array}{r@{\quad,\quad}l}
E (d-z)	& z < d \\
0		& z > d
\end{array} \right.
\end{equation}
and suppose we measure the ion velocity when $z = d$.  In this model the
length scale $d$ is of the order of the axial dimension of a control
electrode, and the electric field $E$ is of the order of the electric
potential applied to a control electrode divided by its axial dimension.
Drag is important when the dimensionless drag parameter $c = \gamma t_d
\gtrsim 1$ where $\gamma = - F_D/(m \dot{z})$ and $t_d$ is the time to reach
$z = d$. Here $t_d$ is the solution to the transcendental equation
\begin{equation}
d = \frac{Q E}{\gamma m} \left( t_d + \frac{e^{-\gamma t_d} - 1}{\gamma}
\right) \ .
\end{equation}
The dimensionless drag coefficient $c$ is plotted as a function of
pressure in \fig{drag-parameters}.  At 70 Pa, $c = 4.9$ so drag plays
an important role in ion shuttling experiments in the cross trap despite
the fact that the trap stability is not affected by damping.

For large damping ($c \gg 1$), $\dot{z}(z = d)$ approaches the terminal
velocity $(Q E)/(\gamma m)$.  The drag coefficient $\gamma \sim
1/C(N_{Kn}) \sim 1/N_{Kn} \sim p$, so $\dot{z}(z = d) \sim 1/p$.  This
is experimentally verified in \sect{movement}.

\section{Planar trap performance}
\label{sec:performance}

When damping forces can be neglected, the Mathieu equations of motion of
a trapped ion depend only on dimensionless parameters. It follows that
the dynamics of macroscopic charged particles, viewed in the appropriate
time scale (a micromotion period), are identical to that of atomic ions,
and can be fully explored without the much more demanding experimental
requirements of trapping atomic ions \cite{Hoffnagle}. We have found
macroscopic charged particle planar traps to be a rapid and accurate
test bed for investigating traps of different geometries (e.g.
three electrode, five electrode) and control electrode layouts (e.g.
segmented center, segmented outer electrodes).  In this section, we
present the results of an experimental investigation of planar trap
performance using this test bed.

\subsection{Ion height above trap substrate}

A planar trap with a segmented center electrode can be used to control
the height of individual ions above the substrate by applying DC
voltages to the electrode segments. One advantage of such control in a
quantum computer is the ability to perform single qubit gates on
different ions using a single stationary laser beam. The laser beam
would have to be parallel to the ion chain but slightly raised or
lowered. By changing the height of individual ions, they can be brought
into the path of the beam to perform single qubit gates.

\begin{figure}
\includegraphics[width=8.2cm]{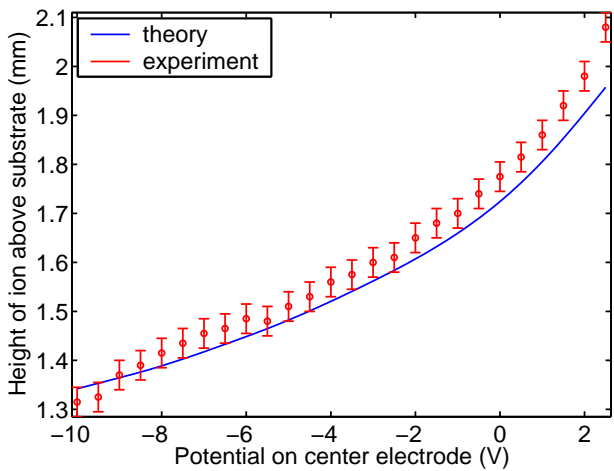}
\caption{\label{fig:ion-height-vs-center-potential}Ion height above the
trap substrate, $r_0$, is plotted as a function of the DC potential on
the center electrode.  The solid line is the expected ion height from
numerical computations of the secular potential as described in
\sect{trap-design} using the measured charge-to-mass ratio of the ion.}
\end{figure}

We measured the height of an ion above the trap substrate 
using a CCD camera mounted on a calibrated translation stage.  The
micromotion of the ion in the $y$ direction causes the images (which have an
exposure time much longer than the period of the RF drive) to show a streak
where the ion is located, so we actually measured the positions of the top
and bottom of the ion motion.  \ffig{ion-height-vs-center-potential} shows
the average of the positions of the top and bottom of the ion motion as a
function of the DC potential on the center electrode.  We measured $Q/m$
for the ion by lowering the RF drive frequency until
the ion became unstable as described in \sect{eoverm}.  The expected
ion height shown in \fig{ion-height-vs-center-potential} is calculated
using this value of $Q/m$ and a numerical computation of the
secular potential.

\begin{figure}
\includegraphics[width=8.2cm]{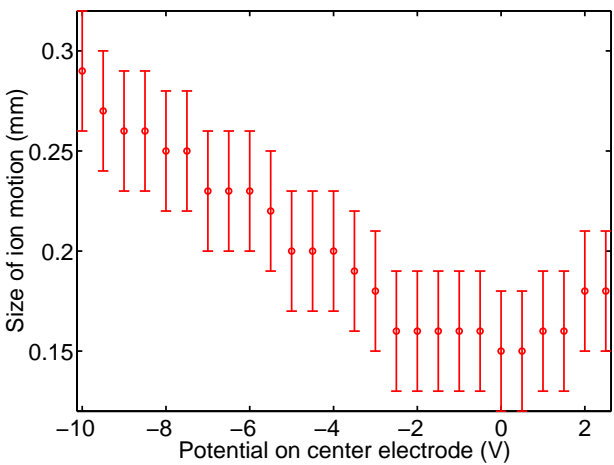}
\caption{\label{fig:ion-size-vs-center-potential}Ion motion
amplitude.  The peak to peak amplitude of the ion motion in the $y$
direction is plotted as a function of the DC potential on the center
electrode.}
\end{figure}

\ffig{ion-size-vs-center-potential} shows the amplitude of the ion motion
as a function of the DC potential on the center electrode.  The
micromotion is minimized when the ion is located at the null of the RF
quadrupole electric field, which occurs when the potential on the center
electrode is zero.  Even without a bias on the center electrode, however,
there is still a substantial amplitude of ion motion because we do not
cool the secular motion of our ions.

\subsection{\label{sec:secular}Secular frequencies and trap geometric factors}

\begin{figure}[h]
\begin{minipage}{0.15\textwidth}
\includegraphics[width=\textwidth]{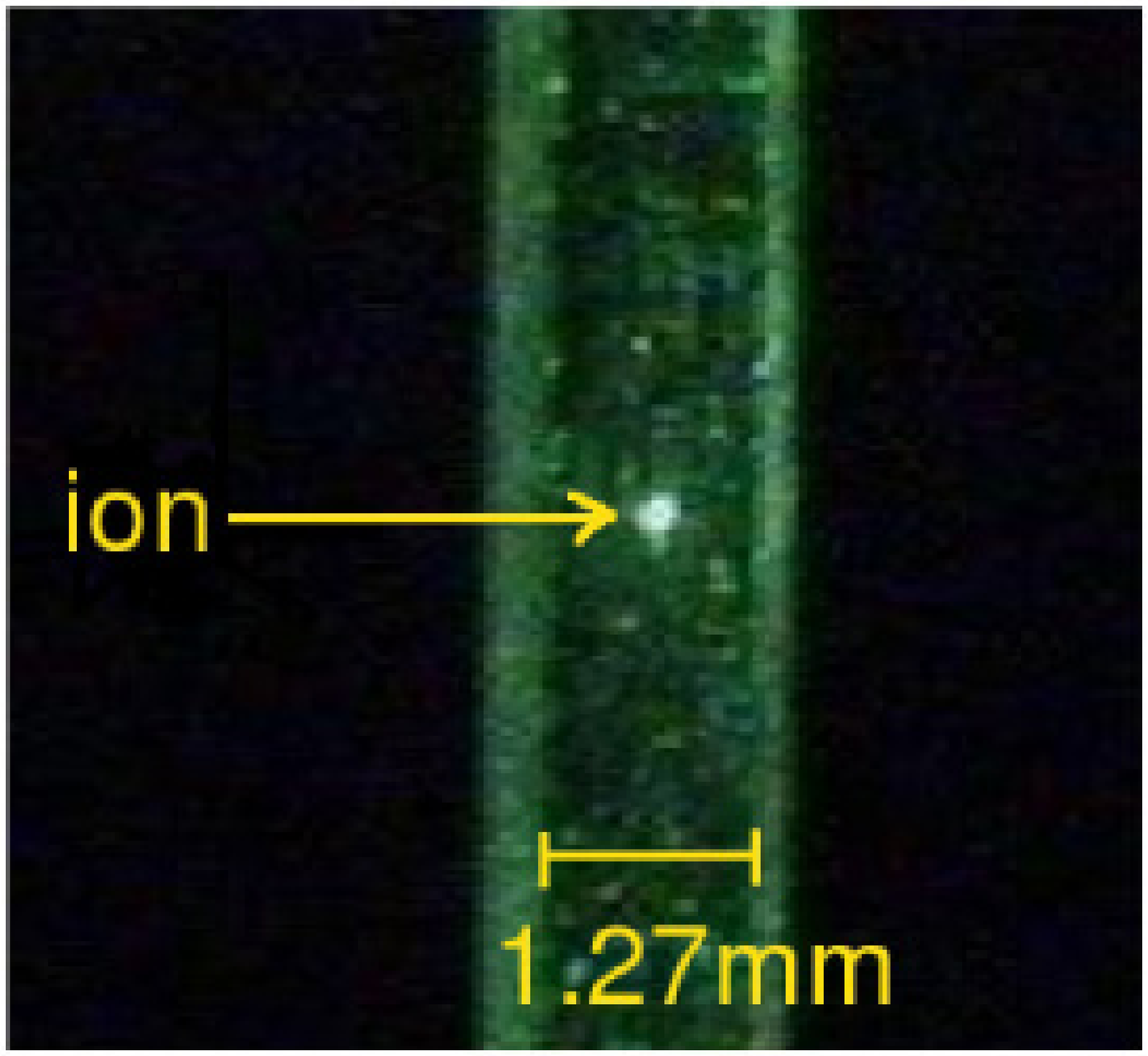}  
(a)
\end{minipage}
\begin{minipage}{0.15\textwidth}
\includegraphics[width=\textwidth]{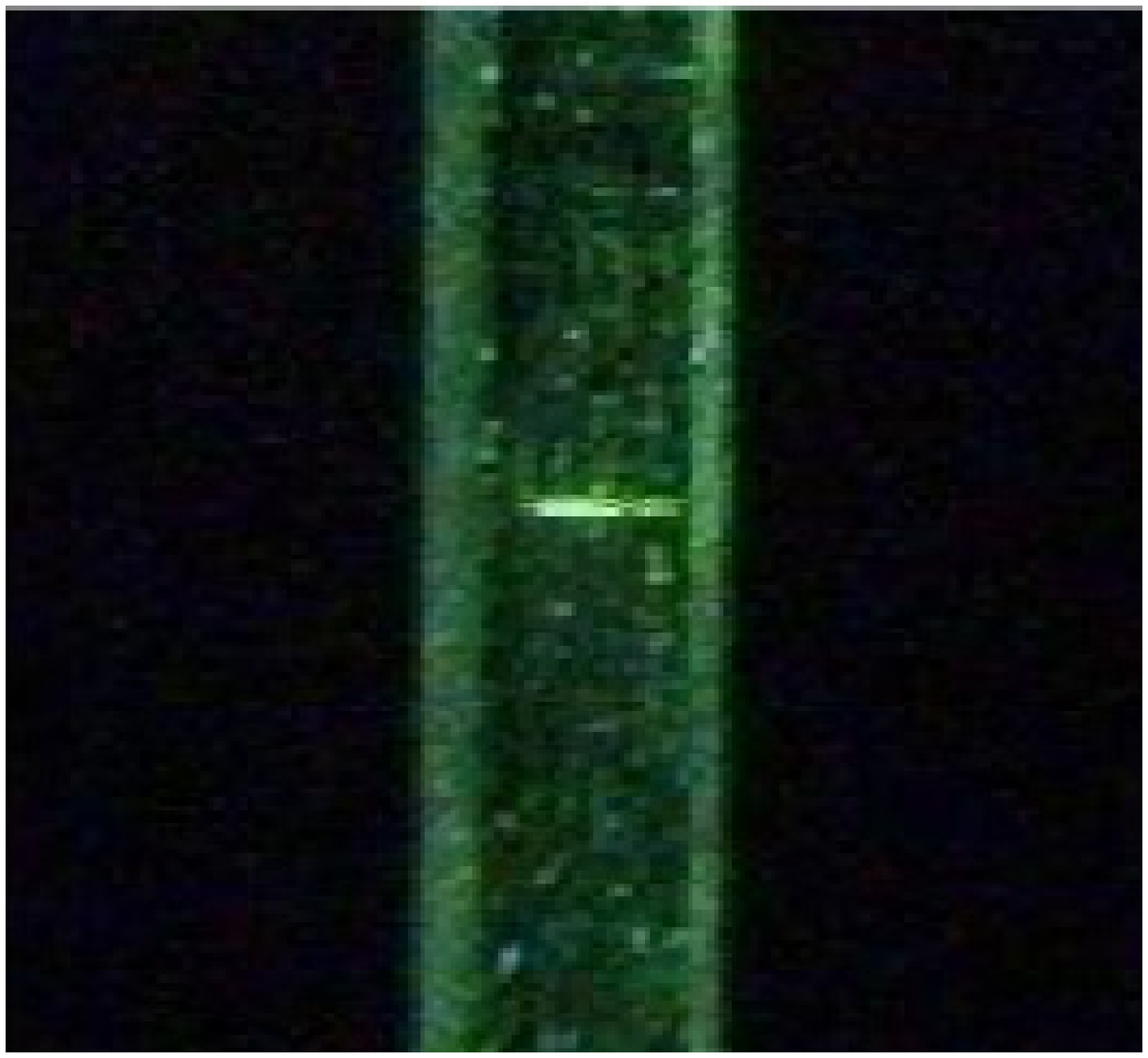} 
(b)
\end{minipage}
\begin{minipage}{0.15\textwidth}
\includegraphics[width=\textwidth]{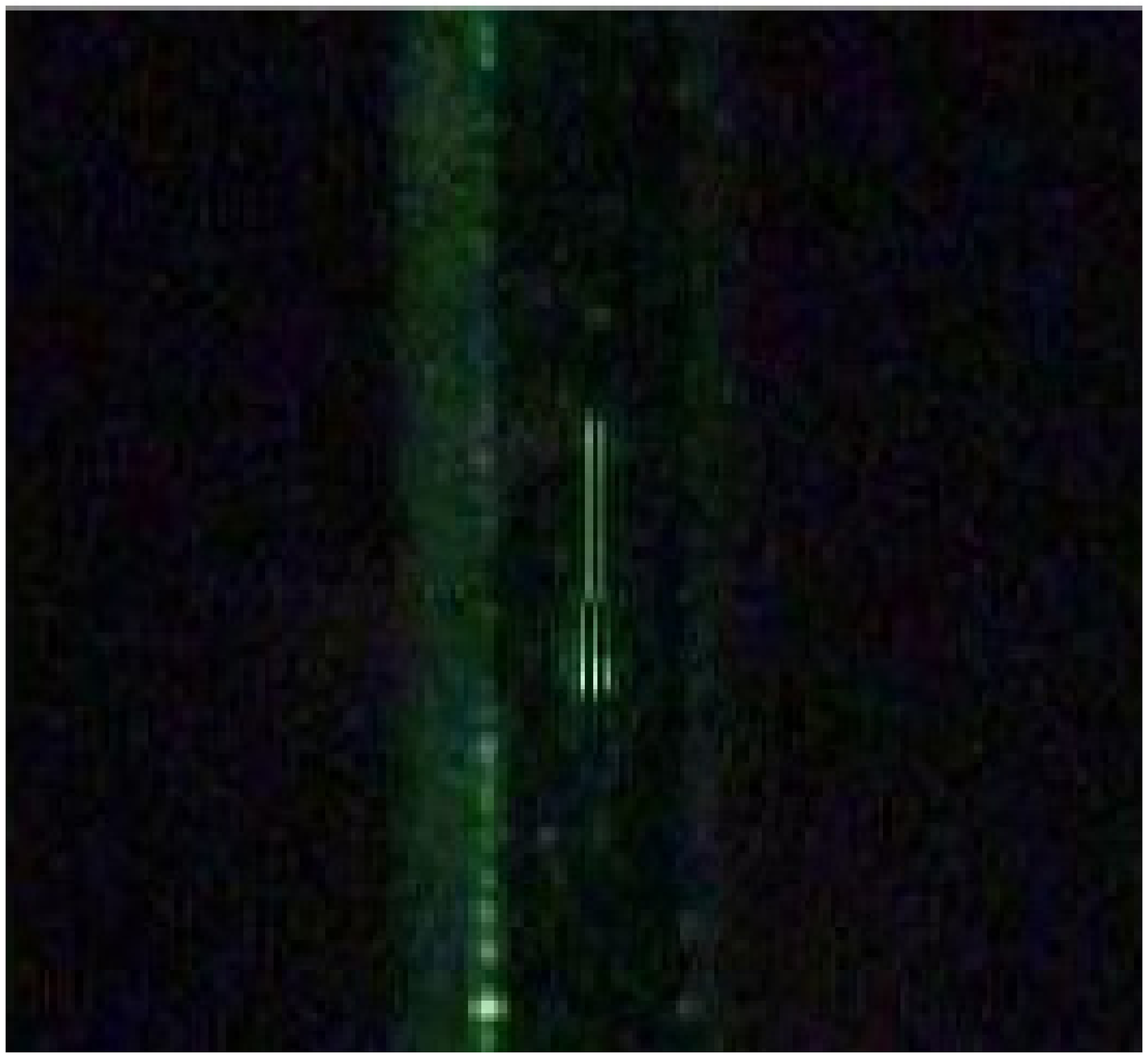}
(c) 
\end{minipage}
\caption{(a) Top view of an ion trajectory from secular resonances with
$V=250$ V and $\Omega = 2\pi \times 2$ kHz. Only the center ground electrode
is seen in these pictures because the grounded top plate electrode masks the
other electrodes from view. The width of the ground electrode is 1.27 mm.
(b) Transverse mode excited at a drive frequency of 288 Hz. (c) Axial mode
excited at 25 Hz.} \label{fig:excitingSecularModes}
\end{figure}

\begin{figure}[h]
\begin{minipage}{0.23\textwidth}
\includegraphics[width=\textwidth]{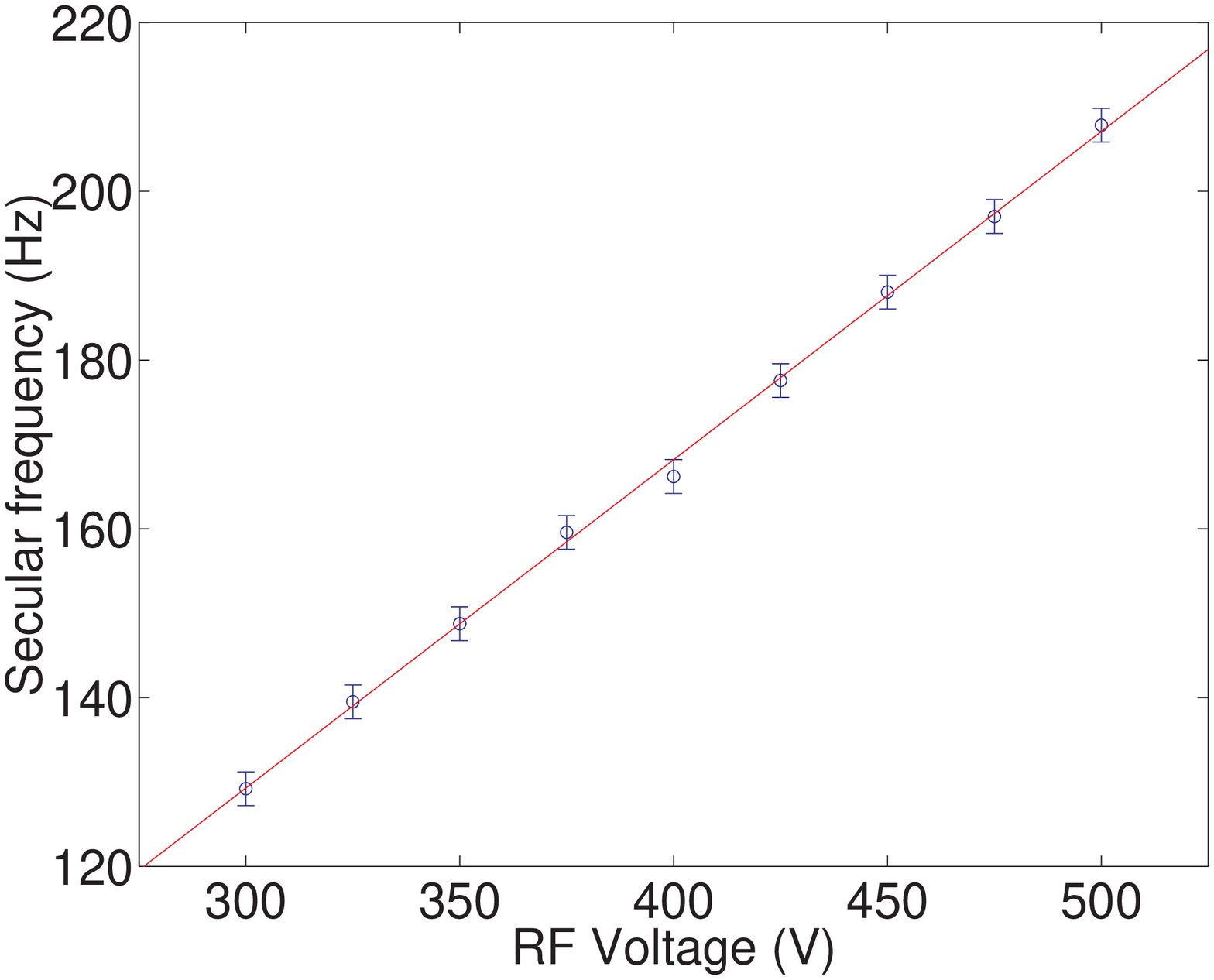}  
(a)
\end{minipage}
\begin{minipage}{0.23\textwidth}
\includegraphics[width=\textwidth]{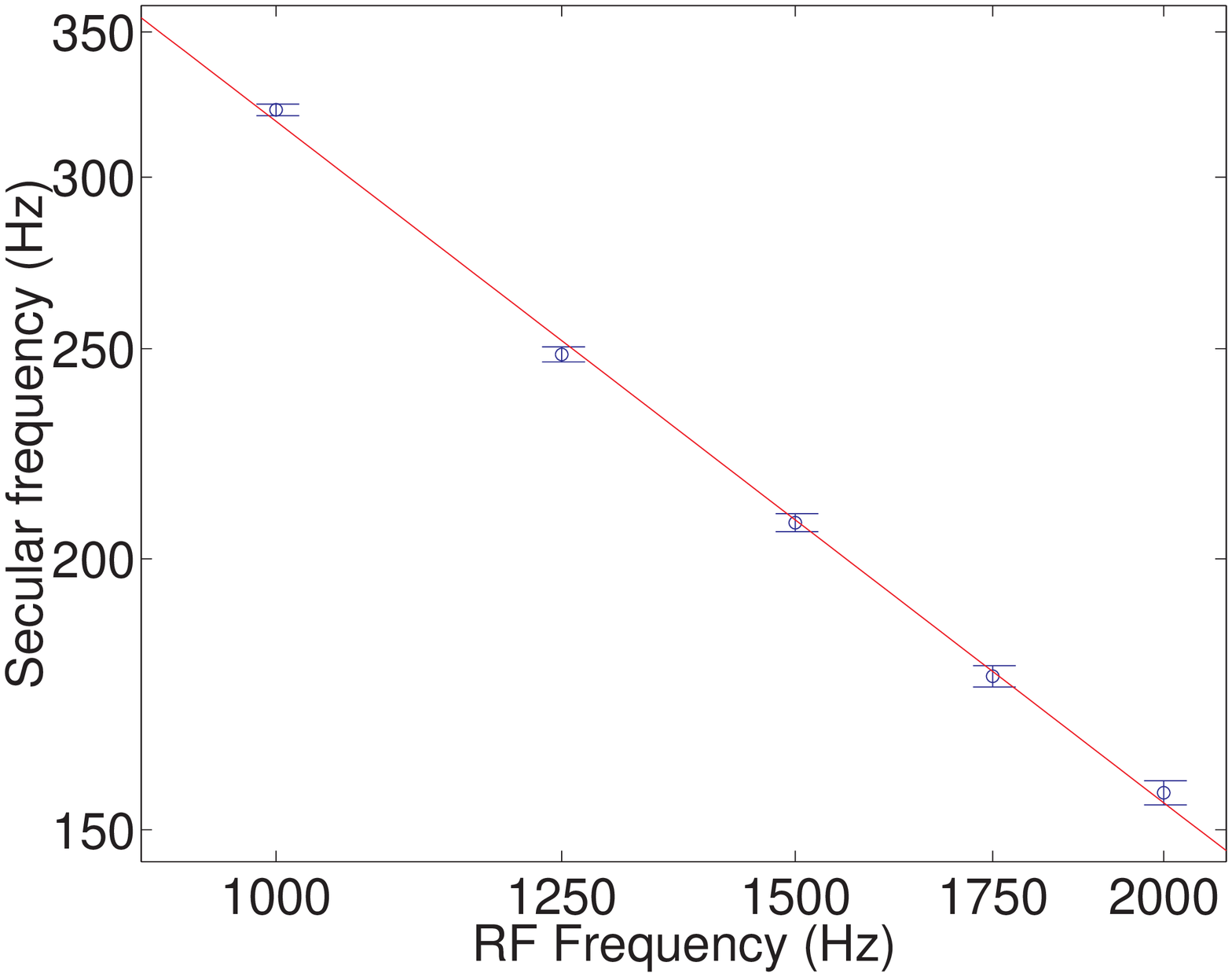} 
(b)
\end{minipage}
\caption{Measurements of the secular frequency versus (a) RF
amplitude at a fixed RF frequency of 1.5 kHz (on a linear scale) and
(b) RF frequency at a fixed RF amplitude of 250 V (on a log-log scale). The
solid lines are fits to \eq{secularfreq} with $\frac{Q}{m}f_i$ as a fit
parameter.} \label{fig:secdependencies}
\end{figure}

We determine the secular frequency of an ion by adding a Fourier
component of frequency $\Omega_{t}$ and amplitude $V_{t}$ to the RF
signal. Typically, $V_{t}$ is between 0.5 V and 2 V. By sweeping
$\Omega_t$, we are able to observe a sharp increase in the size of the
ion trajectory along the direction of the secular mode excited by the
drive signal as seen in \fig{excitingSecularModes}. Although the drive
signal is applied to the RF rods, indirect coupling of the drive to the
axial motion of the ion is sufficient to excite axial modes. The exact
axial secular frequency depends on which trap electrodes are used as
endcaps and what voltage is applied to them, but it is typically between
15 Hz and 30 Hz. The transverse secular frequencies are of more interest
because they depend directly on the trap geometry: the geometric factor
$f_i/r_0^2$ in \eq{secularfreq} depends only on electrode dimensions and
their spacing. We were not able to resolve separate resonant frequencies
for $\omega_x$ and $\omega_y$, which indicates that the ion energy is
much less than the trap depth. \ffig{secdependencies} shows the
dependence of the $x$ secular frequency on $V$ and $\Omega$. Both
dependences behave as expected from the pseudopotential approximation.

To confirm the accuracy of our simulations of pseudopotentials in planar
traps, we measured the ratio of the geometric factor $f_x/r_0^2$ in the
planar trap to that of the loading four rod trap. The large spread in
the $Q/m$ distribution does not allow for the determination of the
geometric factor for each trap separately. However, we can measure the
ratio of the geometric factors of the traps by first measuring
$\omega_x$ of an ion in the planar trap, then applying a sequence of
voltages on the control electrodes to move that ion into the four rod
trap, where $\omega_x$ is measured again. By doing several experiments
of this kind, we determined the geometric factor ratio to be
$0.40\pm0.01$. Numerical simulations of this setup were also performed
using BELA, giving for this ratio the value $0.404$, which is within the
error bar of the experimental result.

\subsection{\label{sec:movement}Ion movement}

Three important ion movement primitives for a multiplexed ion trap
quantum computer are shuttling, turning corners and splitting/joining
two ions~\cite{kielpinski02}.  We have demonstrated these three
operations in planar traps, using our macroscopic charged particle system.

\begin{figure}
\begin{minipage}{0.23\textwidth}
\includegraphics[width=\textwidth]{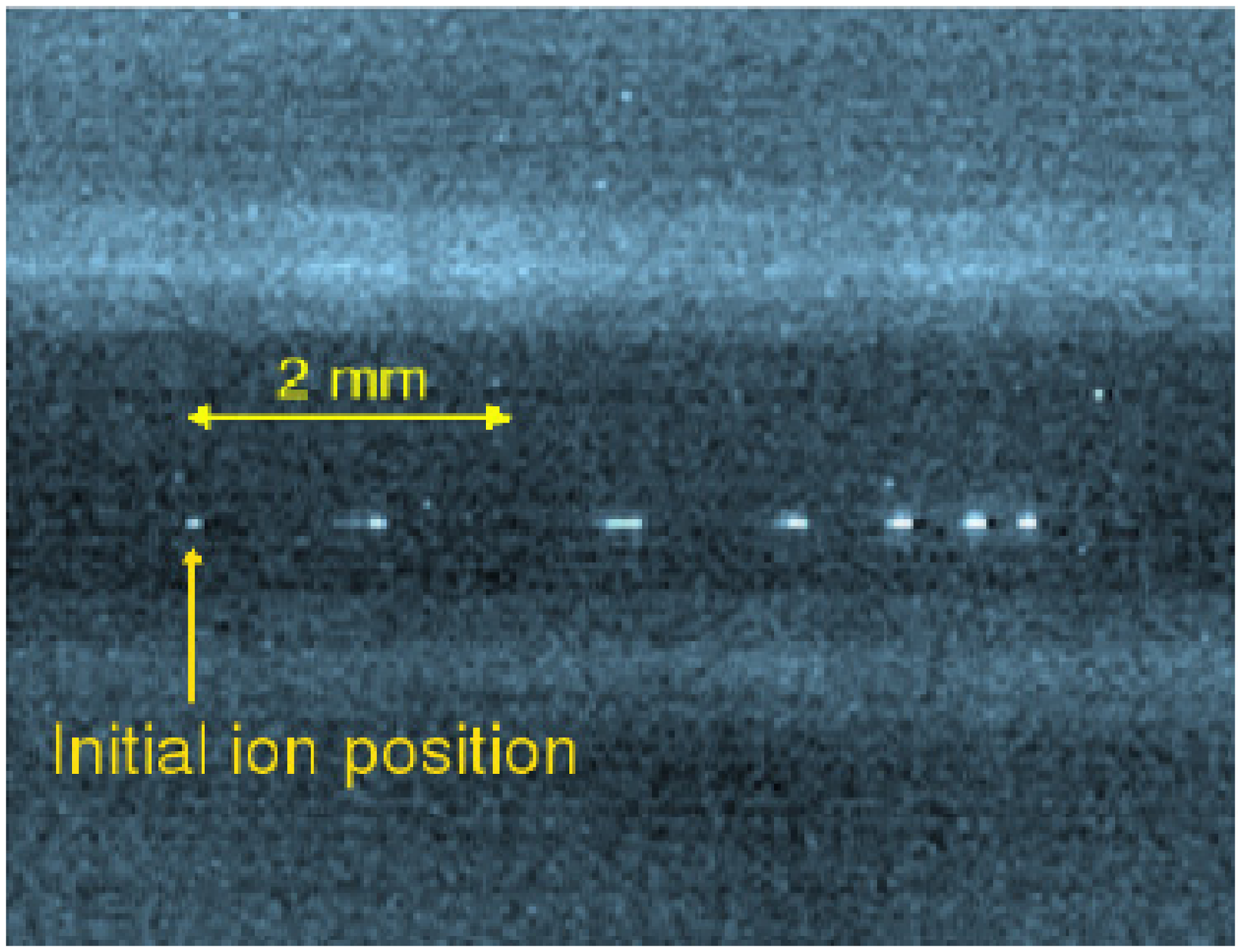}  
(a)
\end{minipage}
\begin{minipage}{0.23\textwidth}
\includegraphics[width=\textwidth]{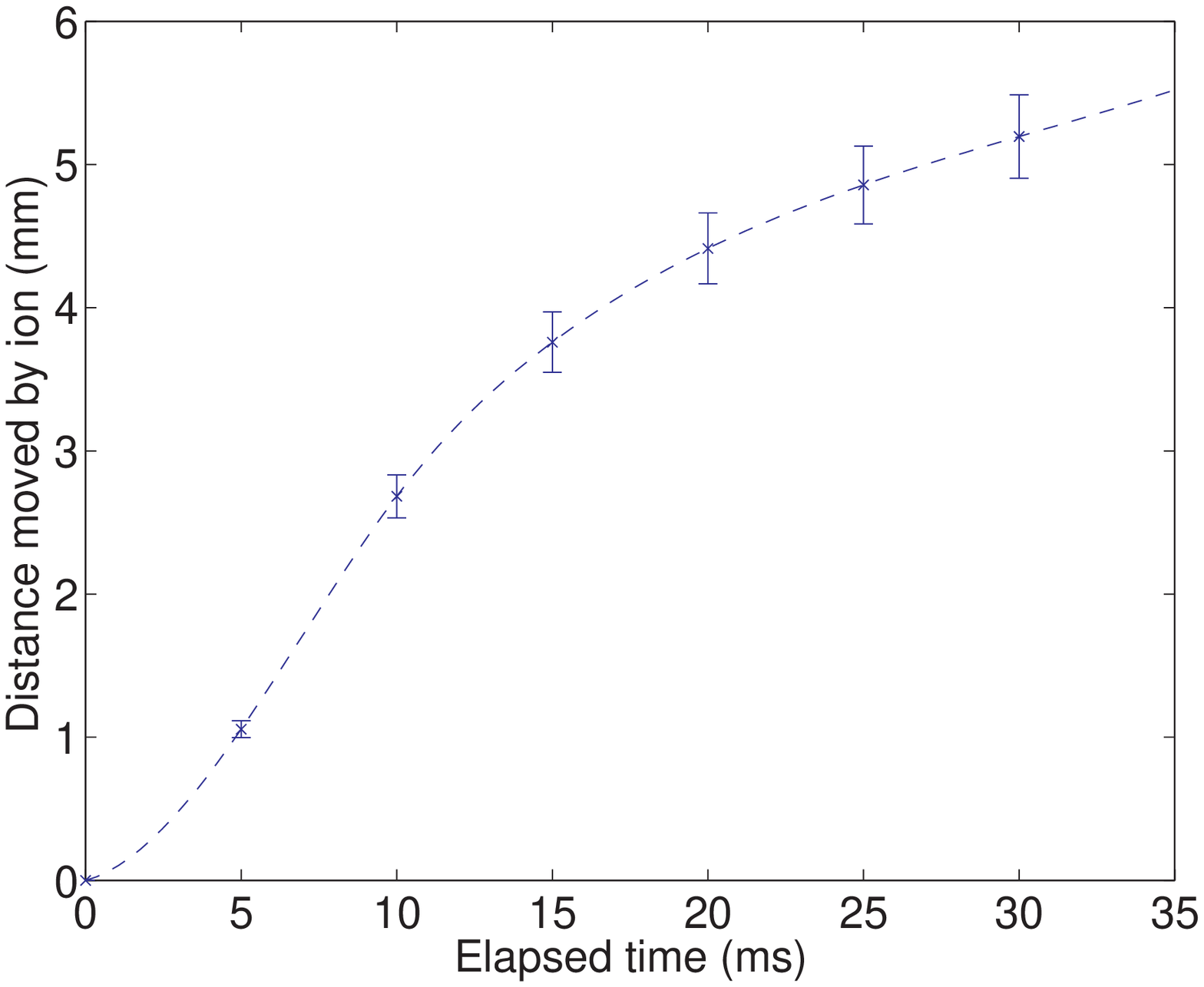} 
(b)
\end{minipage}
\caption{(a) A strobe image showing shuttling of a single
ion, taken by modulating the current to the illuminating laser. The laser is
periodically turned on for 2 ms and off for 5 ms. The ion is moving from
left to right; there is a short acceleration period after which the ion
reaches a maximal velocity and then decelerates. This shuttle was performed
by applying a potential of 5 V to the nearest control electrode that is to
the left of the ion. (b) Distance moved by the ion versus elapsed time,
fitted with a spline to guide the eye.} \label{fig:shuttleStrobe}
\end{figure}

\begin{figure}[t]
\begin{minipage}{0.35\textwidth}
\includegraphics[width=\textwidth]{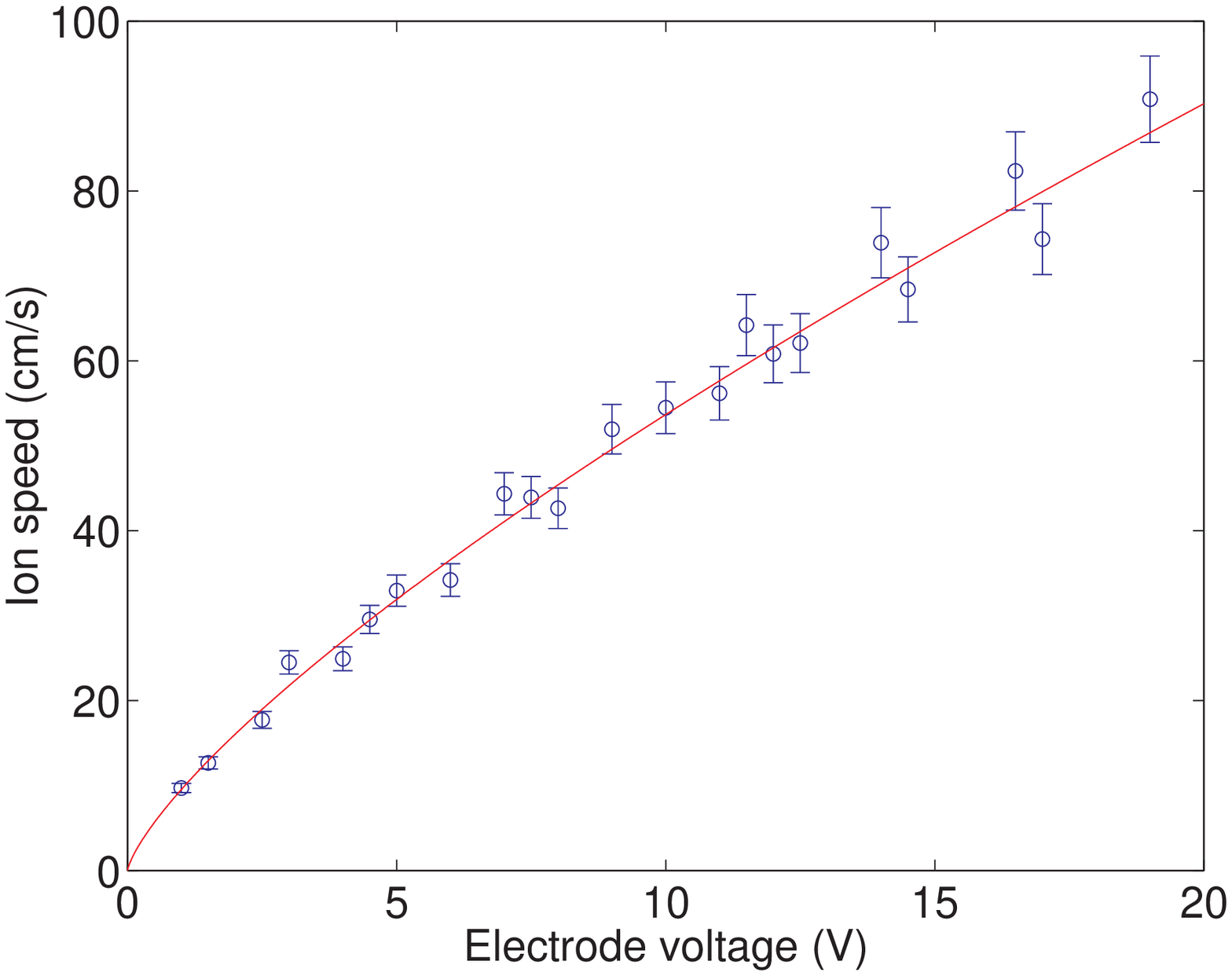}  
\end{minipage}
\begin{minipage}{0.35\textwidth}
\includegraphics[width=\textwidth]{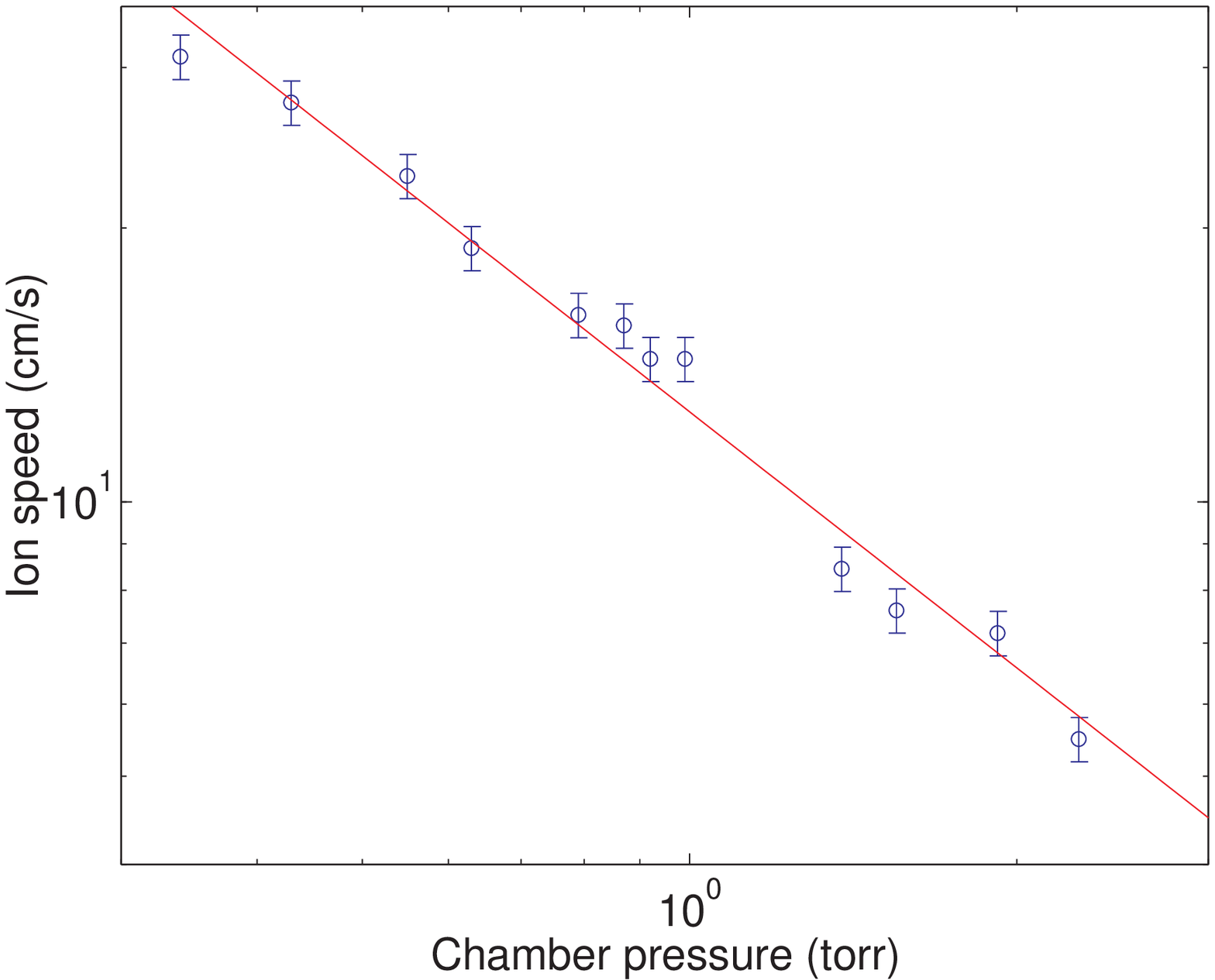} 
\end{minipage}
\caption{The maximum speed $v_{max}$ an ion attains during a shuttling 
operation versus (a) the potential $u_{c}$ applied to the nearest control electrode at
constant pressure (40 Pa) fit to a 3/4 power law ($v_{max} = Au^{3/4}_c$ where $A = 9.55$ cm s$^{-1}$V$^{-\frac{3}{4}}$) (b) the chamber
pressure at constant control electrode voltage (5 V).  Ion speed is
found to be inversely proportional to the chamber pressure.  This
behavior is expected when (1) the mean free path is much greater than
the radius of the particles and (2) $\gamma t_{ac} \gg 1$ where $\gamma$
the damping factor and $t_{ac}$ is the acceleration time of the ion.
When the damping is low enough such that (2) no longer holds, the ion
speed is expected to reach a pressure-independent value.} 
\label{fig:ionspeed}
\end{figure}

\ffig{shuttleStrobe} shows a strobe image of shuttling a microsphere in
the planar trap at 40 Pa (0.34 torr).  The shuttling shown was performed
by raising the potential of the nearest control electrode to the ion. In
this case, the speed of the ion and its final position are determined
both by the electrode potential and the drag force on the ion.
\ffig{ionspeed} shows how the maximum speed the ion achieves during its
motion depends on these two parameters.

We have also shuttled ions using another scheme where the nearest control
electrodes on both sides of an ion are used to create a confining well along
the axial direction. By moving this well along the trap, we had more 
control over the acceleration, speed, and final position of the ion.

Splitting a pair of ions was performed by using the control electrodes
to introduce a potential hill that pushes the ions apart. The ion
spacing is approximately equal to the axial dimension of one control
electrode, so this technique is satisfactory. Typically, an electrode
between the ions is raised to 5 V for separation and lowered back to 0 V
to join them again.  For performing two qubit gates with atomic ions, it
is desirable to bring the ions as close to each other as possible,
because gates are fastest when the frequencies of the two axial normal
modes of the ions are well separated~\cite{Sasura}. In that case, the
ion spacing is likely to be smaller than the electrode spacing, and it
has been shown that it is advantageous to use the DC electrodes to
produce an electric octopole moment for fast separation~\cite{Home}. 

\begin{figure}
\begin{minipage}{0.23\textwidth}
\includegraphics[width=\textwidth]{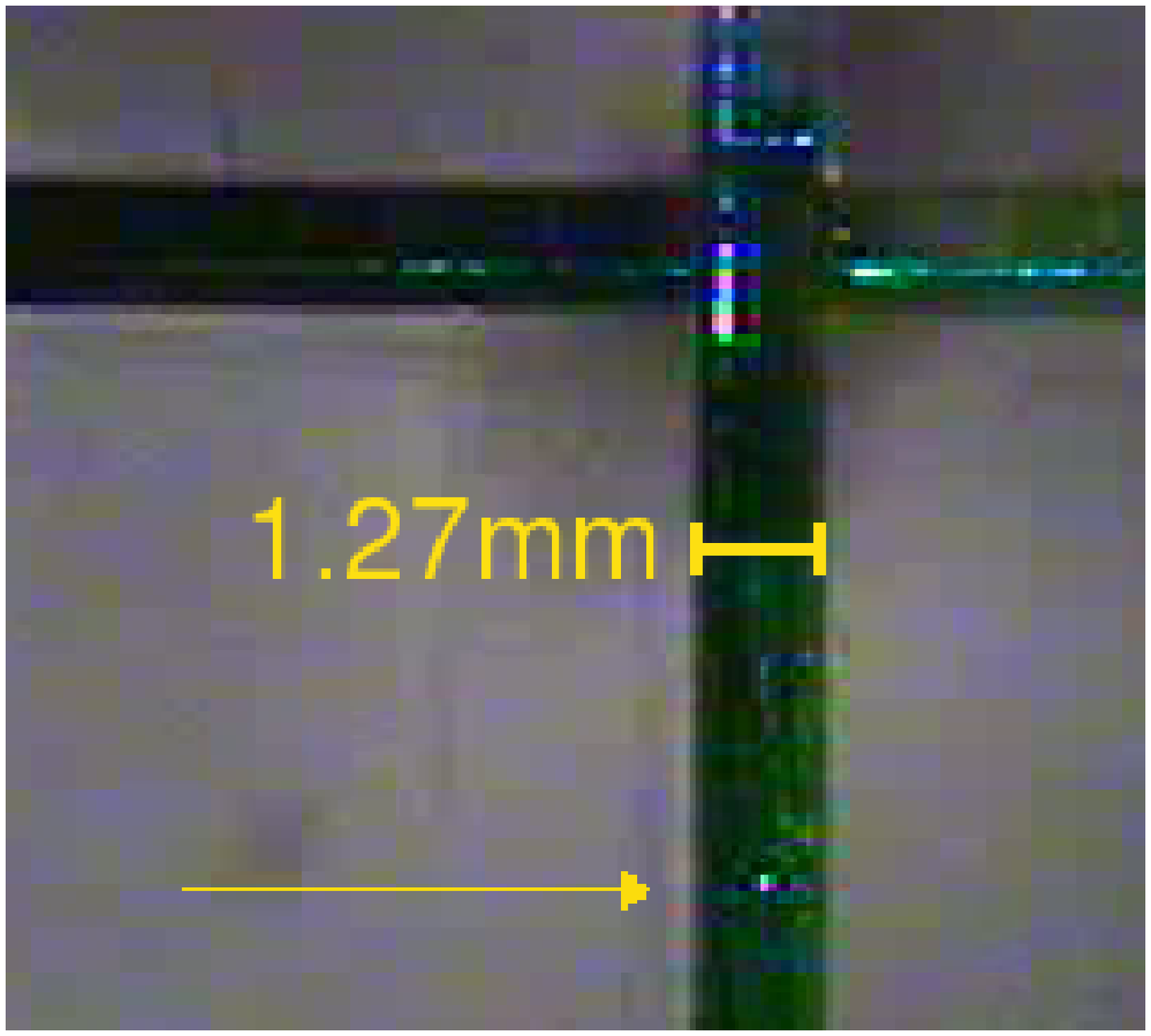}  
(a)
\end{minipage}
\begin{minipage}{0.23\textwidth}
\includegraphics[width=\textwidth]{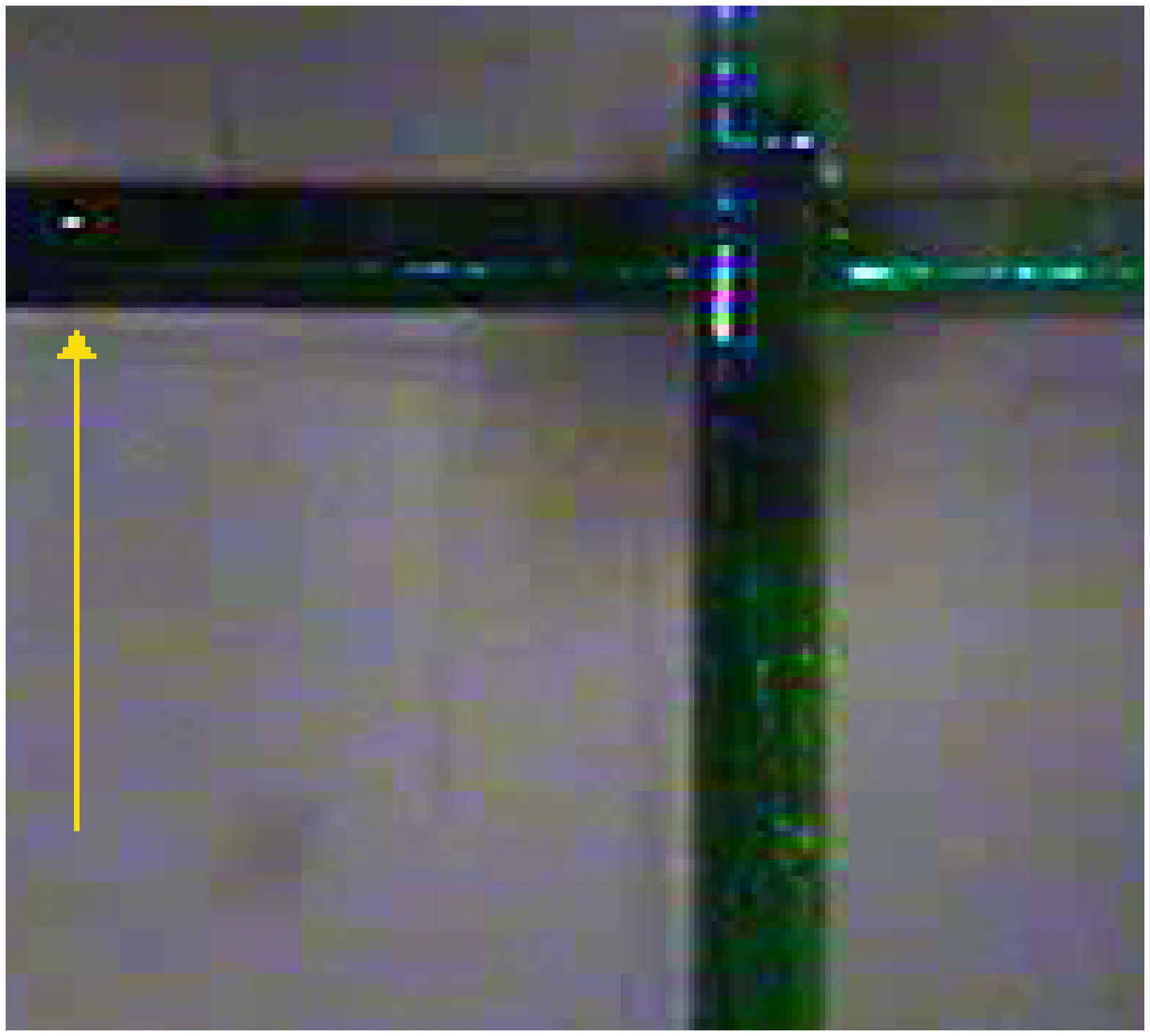} 
(b)
\end{minipage}
\caption{Turning a corner. The arrows point to the
position of the ion (a) before and (b) after turning the corner.  This
operation takes about 50 ms.}
\label{fig:turningcorner}
\end{figure}

\ffig{turningcorner} shows an ion turning a corner in the planar trap.
Simulations of the pseudopotential indicate that there are potential
hills in the RF nodal lines in each of the four arms close to the
intersection, with a point node right at the intersection. To turn a
corner, it is necessary for an ion to overcome the hills in both the
source and destination arms. We have used the following pulse sequence
to do that: (1) lower the center electrode of the destination arm
to -2 V, then (2) raise a control electrode in the source arm to 5 V to
push the ion over the hill in the source arm. We find that the ion has
enough kinetic energy after traversing the first potential barrier to
also overcome the second one.

\section{Variations on trap design}
\label{sec:variations}

Macroscopic charged particles and PCB traps allow us to rapidly test a
variety of planar trap layouts. We have focused on the loading and
transport properties of multi-zone linear traps that are joined at
intersections. In addition to the cross trap described in this paper,
designs under investigation have as many as 100 zones and include
corners, three way junctions, and four way junctions. These elements are
necessary to generate the ion trap geometries that arise in complex
quantum computation geometries \cite{kielpinski02}, particularly those
involved in realizing scalable, fault-tolerant quantum computation
circuits \cite{Svore:04}.

In addition to fault-tolerant quantum computation, ion traps are
promising candidates for quantum simulations. Porras and Cirac have
proposed using the motional modes of coupled ions to simulate
Bose-Hubbard models \cite{PorrasBHM:04}. The planar version of a point
Paul trap is a natural way to implement this scheme for a two
dimensional system. Starting with a plane of RF voltage one can place DC
electrodes at arbitrary positions defining a planar point trap at each
position.  The distance between trapping points controls the strength of
vibrational coupling and the layout of the electrodes can be used to
create both ordered and disordered systems.

\begin{figure}
\includegraphics[width=0.45\textwidth]{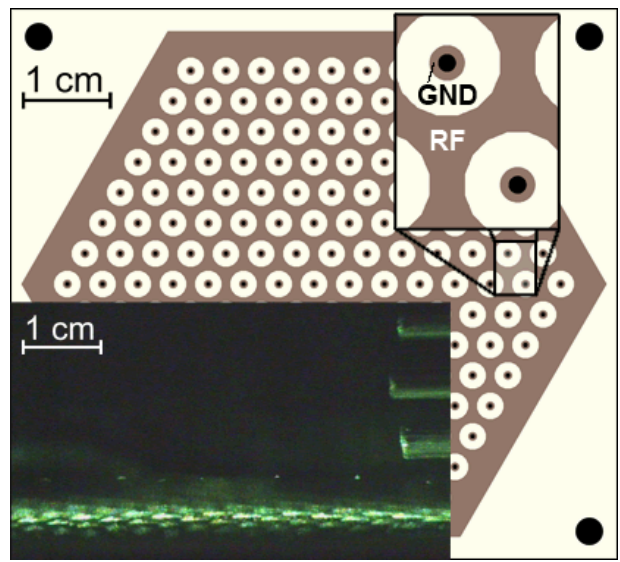}
\caption{Top view of a two dimensional Paul trap array.  Ions can either
be confined to points or free to move along lines, depending on the RF
and ground connections.  The connections shown (top inset) trap ions
above the dots.  Bottom inset: Microspheres are trapped in the 2D array,
and seven are shown illuminated by the laser beam.  In contrast with the
cross trap, the four rod loading stage is out-of-plane, and is visible
above and to the right.}
\label{fig:matrix}
\end{figure}

Using this approach, we have implemented a hexagonal lattice of traps
shown in \fig{matrix}; this lattice successfully trapped macroscopic ions.
For experiments with single atom ions, the lattice spacing needs to be
$\sim$50 $\mu$m to be near interesting phase transitions in
condensed matter systems such as that studied by Porras and
Cirac. Such a trap can be operated as described above with the ions
trapped at the lattice points as shown in \fig{matrix}.  If the RF and
ground electrodes are switched, the ions are trapped along the
honeycomb lines, and applying an offset voltage to the RF on the dots
moves the ions through the lattice. This increased coordination
between ion positions may improve fault-tolerant thresholds.

\section{Conclusions}
\label{sec:conclusions}

We have experimentally demonstrated the planar ion trap design using
macroscopic charged particles as a rapid development test bed. We have
addressed the challenges of ion loading and ion motion given the smaller
relative trap depth of planar traps.

The challenge of loading ions into the planar trap was solved by coupling a
traditional four rod trap to a planar trap.  Ions at high kinetic energy
were first loaded into the linear trap and then offloaded onto the planar
trap. This method allows us to load the planar trap with the efficiency of
loading a traditional linear trap.

Furthermore, we have found that the addition of a charged conductive plane
above the trap can be used to increase the trap depth. In the atomic case
the ions will be initally loaded at a position of high micromotion limiting
the effect of laser cooling. In our design example this micromotion has an
amplitude of $0.04 r_0$ when the potential on the top plane is such that the
trap depth is increased by a factor of 10. However, after initial trapping
the micromotion can be minimized and the cooling maximized simply by
grounding the conductive plane.

Controlled movement of the ions in the planar trap was accomplished
despite the small trap depth. Using increasingly complex traps, we have
been able to perform all the fundamental movement operations required in
a multi-zone architecture: splitting and joining ion chains and moving
ions around corners and through four way intersections. Before this
work, ion movement through a four way intersection was predicted to not
be possible.

Questions remain about the control of ions at low vacuum. Our experiments
were performed at a vacuum where the background pressure no longer effects
trap stability. However, the pressure still contributed a significant drag
term to the linear motion. 

Additionally, we have demonstrated that planar traps can be used to
produce a wide range of geometries. As an example, we have trapped in a
hexagonal lattice.  A similar trap for atomic ions could be used to
simulate two dimensional quantum simulations. One can then controllably
introduce disorder into the system by application of voltages or changes
in trap fabrication.

Planar ion traps are a general tool with many applicatuions in quantum
simulation, quantum computation, and mass spectrometry.  Using modern
two dimensional fabrication techniques, the traps described here can be
reconstructed to be compatible with UHV and atomic ions. Open questions
about the effect of the surface on ion heating can then be addressed.
Planar traps or ``ion chips'' offer an exciting route to study new
atomic ion physics.


\end{document}